# Weak turbulence theory of the non-linear evolution of the ion ring distribution

M. Mithaiwala, L. Rudakov[1], G. Ganguli and C. Crabtree

Plasma Physics Division, Naval Research Laboratory, Washington, DC 20375-5346
[1]Icarus Research Inc., P.O. Box 30780, Bethesda, MD 20824-0780 and
University of Maryand, Departments of Physics and Astronomy, College Park, Maryland 20742

## Abstract

The nonlinear evolution of an ion ring instability in a low-$\beta$ magnetospheric plasma is considered. The evolution of the two-dimensional ring distribution is essentially quasilinear. Ignoring nonlinear processes the time-scale for the quasilinear evolution is the same as for the linear instability $1/\tau_{QL} \sim \gamma_L$. However, when nonlinear processes become important, a new time scale becomes relevant to the wave saturation mechanism. Induced nonlinear scattering of the lower-hybrid waves by plasma electrons is the dominant nonlinearity relevant for plasmas in the inner magnetosphere and typically occurs on the timescale $1/\tau_{NL} \sim \omega(M/m)W/nT$, where $W$ is the wave energy density, $nT$ is the thermal energy density of the background plasma, and $M/m$ is the ion to electron mass ratio, which has the consequence that the wave amplitude saturates at a low level, and the timescale for quasilinear relaxation is extended by orders of magnitude.

# I. Introduction

Ring distributions appear commonly in space plasmas. For example they are observed in association with magnetosonic waves[1] from the plasmapause out to geostationary orbit[2]. Ring distributions form in the magnetospheric ring current due to charge exchange with neutral Hydrogen. They may also form from the earthward convection of ions after substorms when the low energy portion of the ions are lost as they $\vec{E} \times \vec{B}$ drift in the Earth's field.[3,4] Additionally ring distributions are created in a variety of active ionospheric experiments where a beam of neutral atoms is released perpendicular to the magnetic field.[5,6] The photo-ionization of the neutral atoms and subsequent gyration about magnetic field lines generates a velocity ring distribution. They are also observed in the cometary bow shock when a water molecule becomes ionized, and begins gyrating with the solar wind velocity. Space shuttle water releases, which are required to get rid of excess water produced by the fuel cells, produce ion ring distributions.[7]

Ring velocity distributions are highly anisotropic and are unstable with a variety of wave modes.[8,9,10,11,12] In the magnetosphere these distributions are observed along with waves with frequencies ranging from a few harmonics of the proton cyclotron frequency up to the lower hybrid frequency.[3] For the low-$\beta$ magnetosphere, only electrostatic ion-cyclotron waves and lower-hybrid waves are considered here.

The evolution of the ring distribution has previously been considered in both theory and simulation.[13,14,15,16] In one-dimensional hybrid and particle simulations, the saturation of the ring-instability happens due to particle trapping.[15,17] However in two dimensions[14], ignoring non-linear interactions, the instability saturates due to quasilinear relaxation of the ring ions, i.e., the wave energy grows exponentially while simultaneously the waves diffuse the ring ions until the instability has vanished and all waves are absorbed by the ring ions resulting in their thermalization.

In section 2, the basic linear instability theory for the ring distribution is reviewed. In particular, the main electrostatic instabilities relevant for the low-$\beta$ magnetosphere/ionosphere occur at either the ion-cyclotron harmonics or the lower-hybrid frequency depending on whether the ions are magnetized ($\gamma < \Omega_i$, i.e., when the growth rate is smaller than the ion cyclotron frequency) or unmagnetized $\left(\gamma > \Omega_i\right)$ respectively. Additionally the basic equations for the electric field for waves near the lower-hybrid frequency are derived, and it is shown that the electrostatic lower-hybrid waves are the short wavelength limit of the electromagnetic whistler and magnetosonic waves. In section 3, the basic theory for the quasilinear evolution of the ring velocity distribution is described and is shown to be a two-dimensional phenomena. Since the rate for quasilinear evolution, which is governed by the linear instability rate $1/\tau_{QL} \sim \gamma_L$, is very fast compared to the rate of injection of free energy for the instability, e.g., by storm/substorm injections, various nonlinear processes may appear. The main nonlinearities from weak turbulence theory are analyzed. In a low beta plasma the dominant nonlinearity for lower-hybrid waves is nonlinear wave particle scattering, sometimes referred to as nonlinear Landau damping, which is a three-dimensional phenomenon.[18] In section 4, the nonlinear scattering rate $\left(\gamma_{NL}\right)$ is derived. For lower hybrid waves $\gamma_{NL} \sim \omega \dfrac{M}{m} \dfrac{W}{nT}$, which can be quite large even for small $W/nT$. It is shown in section 5 that because of this character, the wave amplitude is maintained at a low amplitude due to nonlinear scattering. A steady state is achieved by the balance of the rates of NL scattering and the loss of the turbulence energy due to ion Landau damping or convection from the region where the ring is localized. This implies that the timescale of the relaxation is not controlled by the linear growth rate, but rather the nonlinear growth rate.

## II. Linear Instability Theory of the Ion-Ring Distribution

Consider an ion-electron plasma with Maxwellian distributions and an ion-ring distribution where the ion-ring species is heavier than the background ions $\Omega_r < \Omega_i$ in a uniform magnetic field $B_0$. The distribution function for ring-ions if the thermal speed $v_{tr}$ of the ring ions is small $V_r/v_{tr} >> 1$, can be approximated as a delta-function distribution

$$f_{\text{ring}}\left(v_\perp, v_\parallel\right) = \frac{1}{\pi^{1/2} v_{tr} V_r} \delta\left(v_\perp - V_r\right) \exp\left(-\frac{v_\parallel^2}{v_{tr}^2}\right), \ (1)$$

where $V_r$ is the ring ion speed. The distribution $f_{\text{j=i,e}}$ of the ion and electron background with number density $n_j$ is taken to be a Maxwellian

$$f_{\text{j=i,e}}\left(v_\perp, v_\parallel\right) = \frac{1}{\pi^{3/2} v_{tj}^3} \exp\left(-\frac{v_\perp^2}{v_{tj}^2}\right) \exp\left(-\frac{v_\parallel^2}{v_{tj}^2}\right), \ (2)$$

where the thermal speed of either the proton(i) or electron(e) species is denoted $v_{tj}$. The thermal speed $v_{tj}$ is related to the temperature of the species $T_j$ using $v_{tj} = \sqrt{2T_j/m_j}$, where $m_j$ is the mass of the species; throughout this article only an isothermal plasma is considered where $T_e = T_i$. The distributions (1) and (2) are normalized such that $2\pi \int f\left(v_\perp, v_\parallel\right) dv_\parallel = f\left(v_\perp\right)$ and $\int v_\perp d v_\perp f\left(v_\perp\right) = 1$.

The main instability to be considered here is the electrostatic lower-hybrid instability. Therefore the frequency range of unstable waves will be restricted to $\Omega_i << \omega << \Omega_e$, though other instabilities such as for low-frequency $\omega << \Omega_i$ inertial Alfven waves are possible[19]. Furthermore, due to the anisotropic nature of the ring distribution, the most unstable wavenumbers occur for $k_\parallel << k_\perp$, where $k_\parallel$ and $k_\perp$ are the components of the wavevector parallel and perpendicular to the magnetic field respectively.

Using the cold plasma model in a dense plasma $\omega_{pe}^2 >> \Omega_e^2$, the current[18] in the frequency range $\Omega_i << \omega << \Omega_e$ is due to the relative drift between the effectively unmagnetized ions and magnetized electrons

$$\vec{j}/n_0 e = -\frac{c\vec{E}_\perp \times \vec{b}}{B_0} - \frac{i\omega}{\Omega_e}\left(1 - \frac{\omega_{LH}^2}{\omega^2}\right)\frac{c\vec{E}_\perp}{B_0} + \frac{ie\vec{E}\cdot\vec{b}}{m\omega} \ . \ (3)$$

Ampere's law $\nabla^2 \vec{A} = -4\pi\vec{j}/c$ in the Coulomb gauge $\nabla\cdot\vec{A} = 0$, along with the electric field equation $\vec{E} = -i\vec{k}\phi + i\omega\vec{A}/c$ where $\phi$ and $\vec{A}$ are the scalar and vector potentials respectively, results in the equations for the electric field

$$E_x = -ik_x\phi\frac{\bar{k}^2}{\bar{k}_\perp^2} \quad E_y = E_x\frac{i\omega}{\Omega_e \bar{k}^2} \quad E_\parallel = -ik_\parallel\phi\frac{\bar{k}^2}{1+\bar{k}^2} \ . \ (4)$$

Taking the divergence of the current $\nabla\cdot\vec{j} = 0$ gives the general dispersion relation for waves with frequency $\Omega_i << \omega << \Omega_e$,

$$\frac{\omega^2}{\Omega_e^2} = \frac{\bar{k}^2}{1+\bar{k}^2}\left(\frac{\omega_{LH}^2}{\Omega_e^2} + \frac{\bar{k}_\parallel^2}{1+\bar{k}_\perp^2}\right), \ (5)$$

where the normalized wavenumber $\bar{k} = kc/\omega_{pe}$ was introduced. $\omega_{pj}^2 = 4\pi n_j e^2/m_j$ is the plasma frequency of species $j$ with charge $q_j = \pm e$, density $n_j$, mass $m_j$, the cyclotron frequency $\Omega_j = eB_0/m_j c$, and $\vec{b} = \vec{B}_0/|B_0|$. From the dispersion relation (5) there are three important limits: when $\bar{k}_\perp >> 1$ and $\bar{k}_\parallel/\bar{k}_\perp << \omega_{LH}/\Omega_e$, $\omega = \omega_{LH} = \sqrt{\Omega_e\Omega_i}$ is the lower hybrid frequency in a dense plasma and both $E_x$ and $E_\parallel$ are electrostatic while the electromagnetic component $E_y$ is small, when $\bar{k}_\perp < 1$ but $\bar{k}_\parallel/\bar{k}_\perp << \omega_{LH}/\Omega_e$, $\omega^2 = k^2 V_A^2$ the dispersion for magnetosonic waves, when $\bar{k}_\perp < 1$ and $\bar{k}_\parallel >> \omega_{LH}/\Omega_e$ $\omega^2 = \bar{k}^2 \bar{k}_\parallel^2 \Omega_e^2$ the dispersion for whistler waves. The electric field (4) shows that all three waves are formally electromagnetic but propagate at different wave normal angles.

Specifically, the unstable waves considered are the lower hybrid waves which are the short wavelength limit, $kc/\omega_{pe} >> 1$, of electromagnetic whistler waves according to (4) and (5). In this limit, the linear dispersion relation can be derived using the electrostatic approximation $\vec{E} = -\vec{\nabla}\varphi$. The general dispersion relation is

$D\left(\omega,\vec{k}\right) = 1 + \sum_j \varepsilon_0^j = 0$, where the sum is taken over all species $j$ where $\varepsilon_0^j$ is

susceptibility of each species.[20,21] If the electron and proton distributions are Maxwellian as in (2) then their susceptibilities may be written[21,22] as

$$\varepsilon_0^e = \frac{2\omega_{pe}^2}{k^2 v_{te}^2}\left(1 + \frac{\omega}{k_\parallel v_{te}} Z\left(\frac{\omega}{k_\parallel v_{te}}\right)\right) + \frac{\omega_{pe}^2}{\Omega_e^2}\frac{k_\perp^2}{k^2}, \quad (6)$$

and

$$\varepsilon_0^i = \frac{2\omega_{pi}^2}{k_\perp^2 v_{ti}^2}\left(1 + \frac{\omega}{k_\perp v_{ti}} Z\left(\frac{\omega}{k_\perp v_{ti}}\right)\right), \quad (7)$$

respectively, where $Z$ is the plasma dispersion function.

The ion susceptibility (7) is equivalent to the susceptibility of a species with a Maxwellian distribution when there is no magnetic field. Formally, the reduction of the background plasma ion susceptibility to its simple unmagnetized form, (7) can be made if $\gamma/\Omega_i + \pi/16 k_\parallel^2 \rho_i^2 > 1$.[11] For flute modes ($k_\parallel = 0$), this reduces to the condition that the growth rate be larger than the cyclotron frequency of the ions $\Omega_{i,r}$, i.e. $\gamma > \Omega_{i,r}$. The procedure for transforming the magnetized susceptibility to its unmagnetized form (7) is treated more fully elsewhere[21,23]

It was demonstrated[24] that when $|\omega - k_\parallel v_\parallel| >> \Omega_{i,r}$, and $k\rho_{i,r} >> 1$ the susceptibility of the ion species for a general distribution can be expressed as

$$\varepsilon_0^j = -\frac{\omega_{pj}^2}{k^2} \int v_\perp dv_\perp \frac{1}{v_\perp} \frac{\partial f_j}{\partial v_\perp} \left(1 - \frac{\omega}{\left(\omega^2 - k_\perp^2 v_\perp^2\right)^{1/2}}\right). \quad (8)$$

When the ring distribution can be expressed as a delta function (2), the susceptibility for ring ions obtained by integrating (8) has a simple form

$$\varepsilon_0^r = -\frac{\omega_{pr}^2 \omega}{\left(\omega^2 - k_\perp^2 V_r^2\right)^{3/2}}. \quad (9)$$

Depending upon the parameters of the ring distribution and background plasma, several types of electrostatic instabilities generated by a ring distribution are possible, and the possibilities are treated in detail elsewhere.[22] For flute modes, $k_\parallel = 0$, if both ion species are effectively unmagnetized, the unstable modes lie near the lower hybrid frequency which is $\omega \approx \sqrt{\Omega_e \Omega_i}$ and the instability involves the ring-ions and the background ions. While if the background ions remain magnetized, the ring distribution is unstable to ion-cyclotron waves.[22]

The electrostatic dispersion relation for unmagnetized ions and magnetized electrons is obtained from the individual susceptibilities (6), (7), and (9). If $k_\parallel / k_\perp << \sqrt{m/M}$, and the electrons and ions are cold $\omega / k_{\parallel,\perp} v_{te,i} >> 1$, then the dispersion relation can be simplified for the background ions and electrons by expanding the Z-function in the large argument limit,

$$D(\omega, k) = 1 + \frac{\omega_{pe}^2}{\Omega_e^2} - \frac{\omega_{pe}^2}{\omega^2}\frac{k_\parallel^2}{k_\perp^2} - \frac{\omega_{pi}^2}{\omega^2} - \frac{\alpha\omega_{pi}^2\omega}{\left(\omega^2 - k_\perp^2 V_r^2\right)^{3/2}} = 0 \text{ , (10)}$$

where it is assumed that $k_\perp \gg k_\parallel$ and $\alpha = \dfrac{n_r/n_e}{M_r/M}$ is the ratio of ring ion number density $n_r$ to electron number density $n_e$ to ring mass $M_r$ to proton mass $M$. Quasineutrality requires $n_e = n_i + n_r$, where $n_i$ is the ion number density. If $k_\parallel/k_\perp \ll \sqrt{m/M}$ then the electron contribution to the dispersion relation (16) represented by the third term on the right hand side of the equality, can be ignored. The subsequent solution, after solving for the real and imaginary parts reveals an instability at the lower hybrid frequency

$$\omega_{LH}{}^2 = \frac{\omega_{pi}^2}{1 + \dfrac{\omega_{pe}^2}{\Omega_e^2}} \text{ , (11)}$$

with growth rate

$$\gamma = \frac{\sin(4\pi/5)}{2}\alpha^{2/5}\omega_{LH} \text{ . (12)}$$

### III. Quasi-linear Theory of the Ring-Velocity Distribution

The quasilinear relaxation of the unmagnetized ring distribution has previously been considered.[13,14] It was shown analytically that the quasilinear relaxation of the unmagnetized ring leads to the stabilization of the ring instability similar to that of the relaxation of a one dimensional beam of fast particles by the Langmuir waves excited by the beam particles. However there is an important difference between the relaxation of a 2-dimensional ring-distribution and a 1-dimensional beam: in the large scale nearly homogeneous magnetized plasma the ring instability generates waves with wave vectors in all directions in the plane normal to the background magnetic field.  The resonant region in 2-dimensional velocity and wave vector space is determined by the condition

$$\omega_k = \vec{k} \cdot \vec{v} = k_\perp v_\perp \cos\varphi \,, \ (13)$$

where the range of unstable wavenumbers lies between $k_1 < |\vec{k}| < k_2$ in the plane perpendicular to the magnetic field and $\varphi$ is the angle between $\vec{k}$ and the velocity vector $\vec{v}$. In 1-dimension the resonant region is simply $\omega = k_\| v_\|$ which allows trapping of resonant particles in the wave potential well.  For the 2-dimensional case the resonant region of velocity space occupies an infinite volume in which $v_\perp > \omega/k_\perp$ just for a single $|k_\perp|$ value which makes trapping not possible due to overlapping wave potential wells. The 1-dimensional relaxation establishes a stationary plateau in the resonant part of the distribution function concurrent with wave spectral energy density. In 2-dimensions the relaxation ends when all waves have been absorbed  by ring and background ions.[13, 14] It is clear that the maximum wavenumber (minimum phase velocity) $k_2$ is determined by the damping of the background ions. The minimum wavenumber is determined from the ring speed $k_1 = \omega / V_r$.

If the ring distribution is also warm, i.e., the cold ring distribution limit of the susceptibility for the ring ions (9) is violated, and the dispersion relation (10) will be

$$D(\omega,k) = 1 + \frac{\omega_{pe}^2}{\Omega_e^2} - \frac{2\omega_{pi}^2}{k_\perp^2 v_{ti}^2}\left(1 + \frac{\omega}{k_\perp v_{ti}} Z\left(\frac{\omega}{k_\perp v_{ti}}\right)\right)$$
$$- \alpha\omega_{pi}^2 \int\limits_{\omega/k_\perp}^{\infty} v_\perp dv_\perp \frac{f_{\text{ring}}}{\left(\omega^2 - k_\perp^2 v_\perp^2\right)^{3/2}} = 0 \quad , (14)$$

with the electron and ion susceptibilities as in (6) and (7) respectively. The last term in (14), which represents the ring distribution, is derived from (8) using integration by parts under the assumption that the ring distribution is narrow and there are no ring particles for $v_\perp \leq \omega / k_\perp$. Then the growth rate due to the ring distribution is

$$\gamma_R = \alpha\frac{\omega_{LH}^4}{k_\perp^2} \int \frac{\partial f_{\text{ring}}}{\partial v_\perp} \frac{dv_\perp}{\sqrt{k_\perp^2 v_\perp^2 - \omega_{LH}^2}} . (15)$$

The stability threshold is due to the competition between the growth of waves from the ring distribution and Landau damping from the background ions. From the dispersion relation, the damping due to the warm thermal background ions can be found similarly to the growth from the ring (15)

$$\gamma_D = \frac{\omega_{LH}^4}{k_\perp^2} \int \frac{\partial f_{\text{back}}}{\partial v_\perp} \frac{dv_\perp}{\sqrt{k_\perp^2 v_\perp^2 - \omega_{LH}^2}} . (16)$$

The thermal background distribution $f_{\text{back}}$ is a Maxwellian distribution as in (2). The total growth rate can be written

$$\gamma_k = \gamma_R + \gamma_D = \frac{\omega_{LH}^4}{k_\perp^2} \int \frac{\partial(f_{\text{back}} + \alpha f_{\text{ring}})}{\partial v_\perp} \frac{dv_\perp}{\sqrt{k_\perp^2 v_\perp^2 - \omega_{LH}^2}} . (17)$$

The set of equations describing the quasilinear evolution of the ring distribution are[13]

$$\frac{\partial f_{ring}}{\partial t} = -\frac{1}{v_\perp}\frac{\partial}{\partial v_\perp}\int\limits_{\omega_{LH}/v_\perp}^{\infty}\frac{\omega_{LH}^4}{n_0 M_r k_\perp}\frac{W_k dk_\perp}{\sqrt{k_\perp^2 v_\perp^2 - \omega_{LH}^2}}\frac{\alpha}{v_\perp}\frac{\partial f_{ring}}{\partial v_\perp}, \text{ (18)}$$

$$W_k = n_0\frac{e^2\left|E_{k,\perp}\right|^2}{2M\omega^2} = W_{k0}\exp\left(2\int\gamma_k dt\right), \text{ (19)}$$

where the growth rate is determined from (15), and the LH wave energy density, $W_k = n_0 e^2\left|E_{k,\perp}\right|^2\big/2M\omega^2$, is due to the electron $E \times b$ drift.

However these equations do not include the evolution of that part of the background distribution that is resonant with the unstable waves. Therefore we extend this model of the quasilinear evolution of the ring distribution to include the background ion distribution by using the total growth rate (17) and including a diffusion equation for the background ions

$$\frac{\partial f_{back}}{\partial t} = -\frac{1}{v_\perp}\frac{\partial}{\partial v_\perp}\int\limits_{\omega_{LH}/v_\perp}^{\infty}\frac{\omega_{LH}^4}{n_0 M k_\perp}\frac{W_k dk_\perp}{\sqrt{k_\perp^2 v_\perp^2 - \omega_{LH}^2}}\frac{1}{v_\perp}\frac{\partial f_{back}}{\partial v_\perp}. \text{ (20)}$$

The set of equations (17)-(20) is valid when the distribution functions $f_{back}$ and $f_{ring}$ as well as wave vector $\vec{k}$ are isotropic in the plane normal to magnetic field lines. Figure 1 shows the numerical solution to the equations for the distributions $f_{back}$ and $f_{ring}$. During the initial growth period, the resonant ring ions transfer their energy to the waves. Eventually the interaction of the ring particles with the waves broadens the ring distribution lowering the growth of the waves and transferring the wave energy back to the ring ions. However during this time the waves also diffuse the background ions creating a non-Maxwellian tail to the distribution. Because the minimum phase velocity is controlled by damping of the background plasma ions, the maximum wavenumber $k_2$ increases during the evolution of the distribution.

As a result of the diffusion of background ions and the creation of a non-Maxwellian tail, the damping from the background plasma ions will be reduced because the background ion distribution flattens. In principle this would lead to a modification of the marginal stability criterion.

The linear and quasilinear analysis of Sections 2 and 3 are valid as long as we ignore the nonlinear processes. Increasing wave amplitude makes nonlinear processes inevitable. The following section will show that the wave amplitudes required for nonlinear interaction between the linear modes are actually small which leads to the early saturation of the ring distribution instability. First, however, the nonlinearities for lower-hybrid waves are examined. [18,25,26]

## IV. Non-Linear Scattering of LH waves

A complete theory for the ring velocity distribution should not only include quasilinear diffusion but also nonlinear wave-wave and wave-particle interaction. For the isothermal low beta magnetosphere, the decay of LH waves to ion-acoustic waves is forbidden and the main nonlinearity is due to nonlinear Landau damping.[18,25,26] For calculating NL effects[*] we will use Vlasov equation in drift approximation for a low-beta plasma[25]

$$\frac{\partial f_e}{\partial t} + v_z \frac{\partial f_e}{\partial z} + c\frac{\vec{E} \times \vec{b}}{B_0}\vec{\nabla}f_e - \vec{\nabla}\left(\frac{c}{\Omega_e B_0}\frac{d\vec{E}_\perp}{dt}f_e\right) - \frac{eE_z}{m}\frac{\partial f_e}{\partial v_z} = 0 . \quad (21)$$

where $f_e = f_e\left(v_\parallel, \vec{r}\, t, \mu \equiv m v_\perp^2/2B_0\right)$, and $\vec{b} = \vec{B}/B_0$. However, for the LH waves the electromagnetic effects are not dominant, $\vec{E} = -\nabla\phi$, and $\vec{b}$ is in the parallel direction. The NL scattering rate follows from imaginary part of the equation of NL dispersion relation

$$k^2\phi_k = \sum_{j=e,i} 4\pi q_j \int dv \left(f_{kj}^{(1)} + f_{kj}^{(2)} + f_{kj}^{(3)}\right), \quad (22)$$

where the nonlinear densities are defined $\delta n_{kj}^{(i)}/n_0 = \int dv f_{kj}^{(i)}$. The superscripts $^{(1)}$, $^{(2)}$, and $^{(3)}$ correspond to the first, second, and third order perturbations of the distribution function $f$ such that $f^{(i)} \sim |\phi|^i$. The linear part of the NL dispersion relation (22), $\varepsilon_k\phi_k = k^2\phi_k - 4\pi e\left(\delta n_{ki}^{(1)} - \delta n_{ke}^{(1)}\right)$, is equivalent to (10).

The first order distribution function is obtained from the last two terms on the left hand side of the drift kinetic equation (21)

---

$$f_k^{(1)} = \frac{e}{m}\frac{i}{\omega - k_\parallel \mathrm{v}_\parallel}E_{k,\parallel}\frac{\partial f^{(0)}}{\partial \mathrm{v}_\parallel} + \frac{e}{m}\frac{i}{\Omega_e^2}k_\perp E_{k,\perp}f^{(0)} \ . \ (23)$$

In a homogeneous plasma, $\vec{\nabla}f_e^{(0)} = 0$, the third term of (21) is zero, thus for the higher order terms

$$f_k^{(n)} = -\frac{1}{\omega - k_\parallel \mathrm{v}_\parallel}\frac{e}{m}\frac{1}{\Omega_e}\sum_{k1+k2=k}\left(\vec{E}_{k1}\times\vec{k}_2\right)_\parallel f_{k2}^{(n-1)} \ . \ (24)$$

However the main NL terms arising from $f_k^{(2)}$ and $f_k^{(3)}$ are obtained only from the third term since $\left(c\vec{E}\times\vec{b}\big/B_0\right)\vec{\nabla}f_e^{(1),(2)} \neq 0$ since the fourth and fifth terms are smaller as $\omega/\Omega_e$ and $k_\parallel/k_\perp$ respectively. Thus the second term on the right hand side of (23) can be neglected from the higher order distribution function (24).

After symmetrizing with respect to $k_1$ and $k_2$, and using the definition for the electric field $E_\parallel = -ik_\parallel \phi_k$, the second order electron distribution function is

$$f_k^{(2)} = \frac{i}{\omega - k_\parallel \mathrm{v}_\parallel}\frac{e^2}{2m^2}\frac{1}{\Omega_e}\frac{\partial f^{(0)}}{\partial \mathrm{v}_\parallel}$$
$$\sum_{k1+k2=k}\left(\vec{k}_1\times\vec{k}_2\right)_\parallel \phi_{k1}\phi_{k2}\left[\frac{k_{2,\parallel}}{\omega_2 - k_{2,\parallel}\mathrm{v}_\parallel} - \frac{k_{1,\parallel}}{\omega_1 - k_{1,\parallel}\mathrm{v}_\parallel}\right] . \ (25)$$

And the third order electron distribution function is

$$f_k^{(3)} = \frac{-1}{\omega - k_\parallel \mathrm{v}_\parallel}\frac{e^2}{2m^2}\frac{1}{\Omega_e}\frac{q}{m}\frac{1}{\Omega_e}\phi_{k1}\phi_{k2}\phi_{k3}\sum_{k1+k2+k3=k}\left(\vec{k}_1\times\left(\vec{k}_2+\vec{k}_3\right)\right)_\parallel\left(\vec{k}_2\times\vec{k}_3\right)_\parallel$$
$$\frac{1}{(\omega_2+\omega_3)-\left(k_{2,\parallel}+k_{3,\parallel}\right)\mathrm{v}_\parallel}\left[\frac{k_{3,\parallel}}{\omega_3-k_{3,\parallel}\mathrm{v}_\parallel}-\frac{k_{2,\parallel}}{\omega_2-k_{2,\parallel}\mathrm{v}_\parallel}\right]\frac{\partial f^{(0)}}{\partial \mathrm{v}_\parallel} \ . \ (26)$$

Integrating the distributions over velocity space,

$$\delta n_e^{(2)}/n_0 = i\frac{e^2}{2m^2}\frac{1}{\Omega_e}\left(\vec{k}_1\times\vec{k}_2\right)_{\parallel}\phi_{k1}\phi_{k2}\left[\frac{k_{2,\parallel}}{\omega_2}-\frac{k_{1,\parallel}}{\omega_1}\right]\int dv\frac{\partial f^{(0)}/\partial v_{\parallel}}{\omega-k_{\parallel}v_{\parallel}},\ (27)$$

$$\delta n_{k1,-k1,k}^{(3)}/n_0 = \frac{-1}{\omega-k_{\parallel}v_{\parallel}}\frac{e^2}{2m^2}\frac{1}{\Omega_e}\frac{e}{m}\frac{1}{\Omega_e}|\phi_{k1}|^2\phi_k\left[\frac{k_{3,\parallel}}{\omega_3-k_{3,\parallel}v_{\parallel}}-\frac{k_{2,\parallel}}{\omega_2-k_{2,\parallel}v_{\parallel}}\right]$$
$$\sum_{k1+k2+k3=k}\left|\left(\vec{k}_1\times\vec{k}\right)_{\parallel}\right|^2\int dv\frac{1}{\left(\omega_{k1}-\omega_k\right)-\left(k_{1,\parallel}-k_{\parallel}\right)v_{\parallel}}\frac{\partial f^{(0)}}{\partial v_{\parallel}}. (28)$$

The nonlinear dispersion relation (22) can be written in a simple form

$$\varepsilon_k\phi_k + \sum_{k1}2\varepsilon_{k-k1,k1}^{(2)}\phi_{k-k1}\phi_{k1} + 2\varepsilon_{k1,-k1,k}^{(3)}\phi_{k1}\phi_{-k1}\phi_k = 0\ ,\ (29)$$

after defining the nonlinear susceptibilities as

$$\varepsilon_{k1,k2}^{(2)} = -\frac{i}{2}\frac{1}{\left(k_1+k_2\right)^2}\omega_{pe}^2\frac{c}{B}\left(\vec{k}_1\times\vec{k}_2\right)_{\parallel}\left[\frac{k_{2,\parallel}}{\omega_2}-\frac{k_{1,\parallel}}{\omega_1}\right]\int dv\frac{\partial f^{(0)}/\partial v_{\parallel}}{\omega-k_{\parallel}v_{\parallel}}, (30)$$

$$\varepsilon_{k1,-k1,k}^{(3)} = \frac{1}{2}\frac{\omega_{pe}^2}{k^2}\frac{c^2}{B^2}\frac{1}{\omega_k}\left[\frac{k_{k,\parallel}}{\omega_k}-\frac{k_{1,\parallel}}{\omega_{k1}}\right]\sum_{k1}\left|\left(\vec{k}_1\times\vec{k}\right)_{\parallel}\right|^2\int dv\frac{1}{\left(\omega_{k1}-\omega_k\right)-\left(k_{1,\parallel}-k_{\parallel}\right)v_{\parallel}}\frac{\partial f^{(0)}}{\partial v_{\parallel}}, (31)$$

using the second and third order density perturbations (27), and (28), respectively. The nonlinear susceptibilities $\varepsilon_{k1,k2}^{(2)}$, $\varepsilon_{k1,k2,k3}^{(3)}$, determine how the fields $\phi_{k1}$, $\phi_{k2}$, and $\phi_{k3}$ couple to produce the nonlinear dispersion relation (29). The superscript $^{(1)}$ is omitted on the linear susceptibility $\varepsilon_k^{(1)} \equiv \varepsilon_k$. The NL dispersion relation (29) determines the dynamics of the potential of the daughter wave $\phi_k$, due to the presence of the pump (mother) wave $\phi_{k1}$ and the low frequency beat wave $\phi_{k-k1}$. The factor of 2 is due to the symmetry between $\varepsilon_{k-k1,k1}^{(2)} = \varepsilon_{k1,k-k1}^{(2)}$ and $\varepsilon_{k1,-k1,k}^{(2)} = \varepsilon_{-k1,k1,k}^{(2)}$.

The low frequency field $\phi_{k-k1}$ is similarly determined from a wave equation for the coupling between the pump wave and the daughter wave $\phi_k$,

$$\varepsilon_{k-k1}\phi_{k-k1} + \sum_{k1} 2\varepsilon_{k,-k1}^{(2)}\phi_k\phi_{-k1} = 0. \ (32)$$

Substituting $\phi_{k-k1}$ from (32) into (29) determines the wave kinetic equation[28]

$$\varepsilon_k\phi_k - \sum_{k1}\frac{4\varepsilon_{k-k1,k1}^{(2)}\varepsilon_{k,-k1}^{(2)}}{\varepsilon_{k-k1}}\phi_{k1}\phi_{-k1}\phi_k + 2\varepsilon_{k1,-k1,k}^{(3)}\phi_{k1}\phi_{-k1}\phi_k = 0. \ (33)$$

Substituting the nonlinear susceptibilities in the wave-kinetic equation (33) yields, in the limit $|\omega_{k1}-\omega_k| << \omega_{k1}$,

$$\varepsilon_k = \sum_{k1}\left|\left(\vec{k_1}\times\vec{k}\right)_\parallel\right|^2\frac{c^2}{B^2}\frac{1}{\omega_k^2}|\phi_{k1}|^2\frac{\omega_{pe}^2}{k^2}\int dv\frac{\left(k_\parallel - k_{1\parallel}\right)\partial f^{(0)}/\partial v_\parallel}{\omega - \omega_1 - (k - k_1)v}\left(\frac{\frac{\omega_{pe}^2}{(k-k_1)^2}\int dv\frac{\left(k_\parallel - k_{1\parallel}\right)\partial f^{(0)}/\partial v_\parallel}{\omega - \omega_1 - (k - k_1)v}}{1 + \varepsilon_i(\omega - \omega_1) + \varepsilon_e(\omega - \omega_1)} - 1\right), \ (34)$$

$$= -\sum_{k1}\left|\left(\vec{k_1}\times\vec{k}\right)_\parallel\right|^2\frac{c^2}{B^2}\frac{1}{\omega_k^2}|\phi_{k1}|^2\frac{(k-k_1)^2}{k^2}\frac{\varepsilon_e(\omega - \omega_1)\left(1 + \varepsilon_i(\omega - \omega_1)\right)}{1 + \varepsilon_i(\omega - \omega_1) + \varepsilon_e(\omega - \omega_1)}$$

where the susceptibilities were defined in (6)-(7). Since $\varepsilon(\omega,k) \approx 0$ near the lower hybrid frequency $\omega \approx \omega_{LH}$, and the growth rate is determined from $\varepsilon(\omega,k) \approx i\gamma_{NL}\,\partial\varepsilon_k/\partial\omega$, where $\partial\varepsilon_k/\partial\omega \approx (2/\omega)\,\omega_{pe}^2/\Omega_e^2$. The LH wave energy density,

$W_{k1} = k_1^2|\phi_{k1}|^2\,\omega\,|\partial\varepsilon/\partial\omega|/8\pi = \left(\omega_{pe}^2/\Omega_e^2\right)|E_{k1}|^2/4\pi$, is primarily due to the energy of the electron $E \times b$ drift. For the subsonic condition that $\Omega_i^2 << (\omega - \omega_1)^2 << \left(\vec{k} - \vec{k_1}\right)^2 C_s^2$ where $C_s^2 \equiv m\mathrm{v}_{te}^2/2M$, $|\varepsilon_i(\omega - \omega_1)| > |\varepsilon_e(\omega - \omega_1)|$ and the NL scattering rate due to thermal electrons is

$$\gamma_{NL} = \frac{\omega_{LH}}{4} \sum_{k1} \frac{\left|\left(\vec{k}_1 \times \vec{k}\right)_{\parallel}\right|^2}{k^2 k_1^2} \frac{M}{m} \frac{W_{k1}}{n_0 T_e} \zeta_e \operatorname{Im}Z(\zeta_e). \quad (35)$$

As already stated, the nonlinear scattering rate is a three-dimensional phenomenon.[18] It was shown in section 2 that lower-hybrid waves are the electrostatic limit of a more general completely electromagnetic dispersion relation. The nonlinear scattering rate (35) for lower-hybrid to lower-hybrid scattering was recently extended include scattering from lower-hybrid to whistler/magnetosonic waves.[18] The resulting scattering rate is

$$\gamma_{NL} \sim \frac{\omega_{pe}^2}{\omega_k} \frac{\bar{k}^2}{1+\bar{k}^2} \sum_{k1} \frac{\left|\left(\vec{k}_1 \times \vec{k}\right)_{\parallel}\right|^2}{k^2 k_1^2} \frac{\bar{k}_1^2}{1+\bar{k}_1^2} \frac{\zeta_e \operatorname{Im}Z(\zeta_e)}{\left[1+\left(\omega_{k1}-\omega_k\right)^2 \Big/ \left(\bar{k}_1 - \bar{k}\right)^2 C_s^2\right]^2} \frac{\left|E_{k1}\right|^2}{4\pi n_0 T_e}. \quad (36)$$

This can be understood as a generalization of the scattering rate (35) where instead of the electrostatic field $E_{\parallel} = -ik_{\parallel}\phi_k$, the fully electromagnetic parallel electric field $E_{\parallel} = -ik_{\parallel}\phi\,\bar{k}^2 \Big/ \left(1+\bar{k}^2\right)$ from (4) is used. NL scattering rates (35) and (36) correspond to the broadband turbulence $\delta\omega >> \gamma_{NL}$ case. The detailed analysis of the parametric decay of the monochromatic LH wave $\delta\omega << \gamma_{NL}$ in a non-isothermal plasma $T_e >> T_i$ when the beat wave $\left(\omega_{k1} - \omega_k\right)$ meets the resonance condition with and ion sound wave resulting in vanishing real parts of the linear dispersion relation (i.e., $\operatorname{Real}(1+\varepsilon_i+\varepsilon_e)=0$) in the denominator of (34) has been analyzed elsewhere.[26]

Physically, the induced scattering from the electrostatic lower-hybrid waves to the electromagnetic whistler waves can be understood by examining the dispersion relation (5), which is shown in Figure 2. In each scattering event, the frequency decreases slightly and the waves follow a path of near constant frequency. On the other hand the wave vector changes substantially. For $\bar{k} \sim 1$, the waves can acquire an electromagnetic component and may convect away from the region where the ring distribution (and hence the turbulence) is localized. Assuming that the characteristic dimension of the region of localization is $L$, the wave convection time is $\tau_{conv} \sim L / V_A$ where the group velocity is approximately $V_A$. This represents an energy loss mechanism that is not accounted for in the quasilinear picture discussed in Sec. 3.

For broadband turbulence $\left( \delta\omega \sim \omega >> kC_s \right)$, the rate of a wave $\left( \vec{k}_1, \omega_{k1} \right)$ to scatter into a wave $\left( \vec{k}, \omega_k \right)$ is obtained by summing the contribution from all waves within the frequency interval $\omega_k + \Delta\omega$ where $\Delta\omega \sim \left| \vec{k}_1 - \vec{k} \right| C_s \sim \left| \vec{k}_1 + \vec{k} \right| \beta_e^{1/2} \omega_{LH}$. After summation this leads to an estimate for the rate

$$\gamma_{NL} \sim \omega_{LH} \left| \bar{k}_1 + \bar{k} \right| \beta_e^{1/2} \frac{\bar{k}^2}{1 + \bar{k}^2} \frac{\bar{k}_1^2}{1 + \bar{k}_1^2} \frac{M}{m} \frac{W_{k1}}{n_0 T_e} \quad (37).$$

For unstable ring distribution the nonlinear scattering rate (35)-(37), can be estimated by noting that the unstable wave vectors are $\left| \bar{k}_1 - \bar{k} \right| \sim \omega_{LH} / V_r$, so that $\left| \bar{k}_1 + \bar{k} \right| \beta_e^{1/2} \sim C_s / V_r$ which makes $\gamma_{NL} \approx \omega_{LH} \frac{C_s}{V_r} \frac{M}{m} \frac{W_{k1}}{n_0 T_e}$. This rate can be large for just small wave amplitudes because of the mass ratio $M/m$, and may be comparable to the linear instability rate itself (12). Therefore the effect of nonlinear scattering on the quasilinear theory from Section 3 must be considered.

# V. Modified Quasilinear Evolution due to Nonlinear Scattering

The equation for the growth of wave energy from the linear instability was stated in (19) where the wave energy grows exponentially. In quasilinear theory the ring particles diffuse so as to limit the growth of wave energy. In this description all the waves are resonant with the ring particles. However the inclusion of nonlinear scattering will pump energy from wavenumbers $k_1$ to $k$ according to (35)-(37), and a new set of waves will be introduced which are not resonant with the ring particles. Thus (19) will be modified to include nonlinear scattering,

$$\frac{\partial W_{k1}}{\partial t} = 2\left(\gamma_{k1}^{L} - \gamma_{k1}^{NL}\left(W_{k}\right)\right)W_{k1}, \text{ (38)}$$

and this will also limit the growth of the resonant wave energy. The nonlinear growth rate $\gamma_{k1}^{NL}\left(W_{k}\right)$ depends on the amplitude of waves $W_{k}$, which are not resonant with the ring ions, and will not be absorbed according to the quasilinear description of Section 3. The amplitude $W_{k}$ is similarly determined by the scattering rate $\gamma_{k}^{NL}\left(W_{k1}\right)$ and the loss of energy $\gamma_{loss}$. The rate of loss may for example be due to nonlinear scattering of electrostatic lower hybrid waves to electromagnetic whistlers that may convect away as discussed in Section 4.

$$\frac{\partial W_{k}}{\partial t} = 2\left(\gamma_{k}^{NL}\left(W_{k1}\right) - \gamma_{loss}\right)W_{k}. \text{ (39)}$$

Because of the energy loss, only a small fraction of the energy extracted from the ring distribution by the instability is reabsorbed by the ring ions contrary to the standard quasilinear theory.

In a steady state, the wave saturation from (38) will occur approximately for $\gamma_{k1}^L = \gamma_{k1}^{NL}$. Estimating the linear growth (12) as $\gamma_{k1}^L \sim \omega_{LH}\left((n_r/n_0)\big/\left(M_r/M_0\right)\right)^{2/5}$, and using the nonlinear growth rate $\gamma_{NL} \approx \omega_{LH}\dfrac{C_s}{V_r}\dfrac{M}{m}\dfrac{W_{k1}}{n_0 T_e}$, the saturated wave energy is

$$\frac{W_{k1}}{n_0 T_e} \approx \frac{m}{M}\frac{V_r}{C_s}\left(\frac{n_r/n_0}{M_r/M}\right)^{2/5}. \ (40)$$

Since all the energy extracted from the ring distribution is not reabsorbed, the time scale for quasilinear relaxation of the ring will be slower and is estimated as

$$1/\tau_{NL} \sim \gamma_L \frac{W_{k1}}{E_{ring}} \sim \omega_{LH}\left(\frac{n_r/n_0}{M_r/M}\right)^{4/5}\frac{m}{M}\frac{n_0 m v_{te}^2}{n_r M_r V_r^2}\frac{V_r}{C_s} \sim \omega_{LH}\left(\frac{n_r/n_0}{M_r/M}\right)^{-1/5}\left(\frac{m}{M_r}\right)\frac{T_e}{M_r V_r C_s}. \quad (41)$$

The amount of energy in the ring is $E_{ring} \approx n_r M_r V_r^2$. For the solar-wind comet interaction[29] this energy may become available for wave generation by photoionization of water molecules from the comet. The rate (41) is mostly independent of the small ring density since $\left(n_r/n_0\right)^{1/5} \sim 1$. The typical parameters of the ring distribution and background plasma are: the ring mass, $M_r/M \sim 18$ for the water molecule, the ring velocity $V_r \sim 500 km/s$, $T_e \sim 10 eV$, and $C_s \sim 30 km/s$ in the solar wind. For these parameters the rate of quasilinear relaxation of the proton ring distribution is rather small $1/\tau_{NL} \sim 3\cdot 10^{-4}\,\omega_{LH}$. Thus the timescale $\tau_{NL}$ (41) is at least $10^3$ times longer than when nonlinear scattering is not included. Consequently the numerical simulation of this nonlinear phenomenon is highly CPU time intensive.

Thus the inclusion of nonlinear scattering has modified the theory of the ring quasilinear relaxation in two important ways. The quasilinear theory must be extended to three-dimensions. Nonlinear Landau resonance is lost in a 2D PIC simulation if background magnetic field $B_0$ is in the plane of simulation. Furthermore, in simulations with $B_0$ in the simulation plane the scattering diminishes strongly[18] because $\left(\vec{k}_1 \times \vec{k}\right)_{\parallel}^2 = 0$. And finally the quasilinear evolution timescale of the ring distribution instability is extended by orders of magnitude.

## VI. Conclusion

The quasi-electrostatic instability of the ion ring distribution for the low-$\beta$ magnetosphere that was considered here is for lower-hybrid waves. Previous studies assumed that although the ring distribution is two-dimensional the analysis can be done in one dimension. Simulations done in this fashion showed that the instability saturates by trapping.[15, 17] In two-dimensions however, the stabilization of the ring instability is essentially quasilinear, i.e., the wave amplitude grows exponentially while simultaneously diffusing the ring ions as to limit the instability. While ignoring nonlinear processes the time-scale for the quasilinear evolution is the same as for the linear instability $1/\tau_{QL} \sim \gamma_L$. In nature the free energy for the instability is supplied on a timescale much longer than the linear timescale, and the analysis for the quasilinear evolution should include nonlinearities.

For lower-hybrid waves, the nonlinear scattering or nonlinear Landau damping, with the rate $\gamma_{NL} \sim \omega \dfrac{M}{m} \dfrac{W}{nT}$, is a three-dimensional phenomenon. The scattering rate for lower hybrid waves can be extended to the whistler/magnetosonic wavevector range. The PIC simulation[30] by McClements et al. of a ring distribution shows that at the midpoint of the simulation when the initially rapidly rising wave energy begins to slow down, approximately 5% of the ring energy is transferred to electron heating (at the end of the simulation it is 10%) while amplitude of waves is 0.1% of the ring energy. This is contrary to quasilinear theory where the energy lost by the ring in the initial unstable phase is about the energy of the waves, i.e. the energy of the waves should be about 5% of the ring energy as well. Thus there is more than an order of magnitude discrepancy between the energy lost by the ring and energy density of waves, which is in qualitative agreement with the nonlinear theory presented where wave energy density saturates at a relatively low level due to nonlinear processes (nonlinear scattering in our case), and is redistributed in k-space to be ultimately absorbed by plasma. In reality, after scattering the newly born electromagnetic waves could convect away from the region of creation. This possibility that the energy is not available for reabsorption by the ring ions leads to a modification of the ring quasilinear evolution. In quasi-steady state the amplitude of waves is estimated as $W_{k1}/n_0 T_e \approx m/M$. Subsequently the quasilinear relaxation timescale of the unstable ring distribution is extended by orders of magnitude.

**Acknowledgements**

This work is supported by the Office of Naval Research.

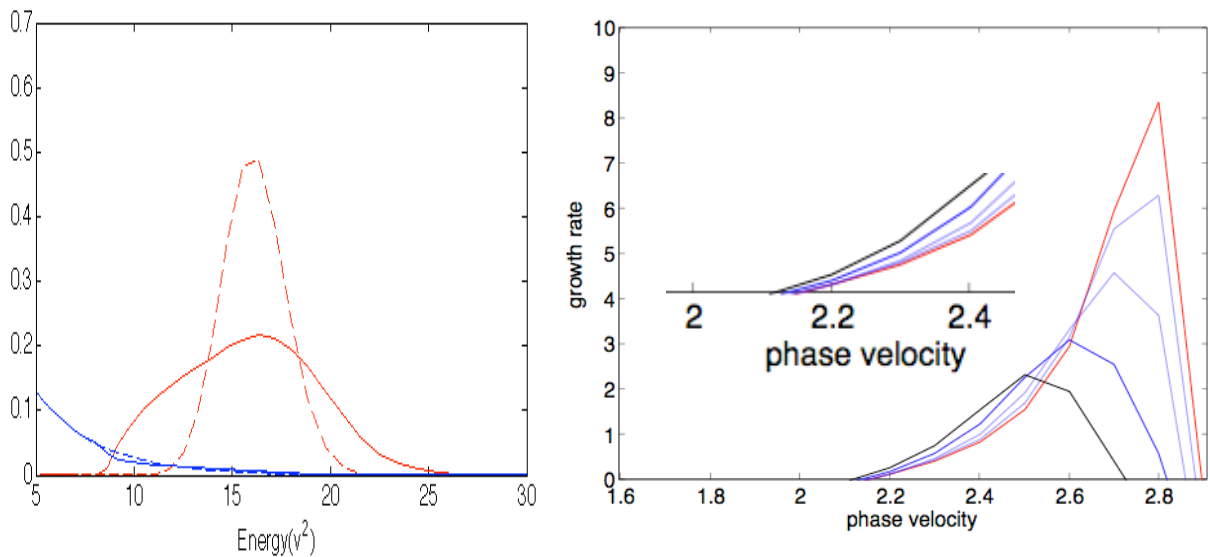

Figure 1. **a)** The initial ring distribution (shown by the red dashed line) and the initial background ion distribution (shown by the dashed blue line). The solid line shows the distributions at a later time according to (19). The wave growth at low energy ($v^2$) leads to diffusion of the background ions and the creation of a non-Maxwellian tail. **b)** The instantaneous growth rate as (18). The creation of the non-Maxwellian tail leads to a reduction in the damping rate at large $k_\perp$ (small phase velocity). Thus at later times, the growth occurs at larger $k_\perp$.

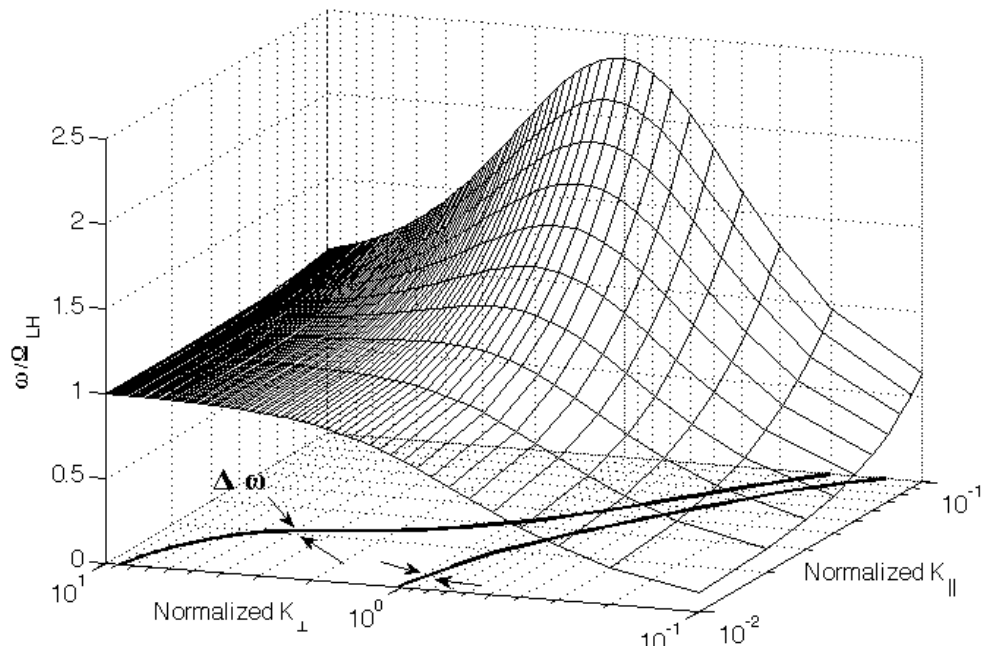

Figure 2. Dispersion surface for intermediate frequency range waves $\Omega_H \ll \omega \ll \Omega_e$ with dispersion relation (5). In each scattering, the waves follow a path of near constant frequency, and can scatter from lower-hybrid waves to whistler waves.